# Magnesium for dynamic nanoplasmonics

Xiaoyang Duan[†,‡] and Na Liu*[,†,‡]

[†]Max Planck Institute for Intelligent Systems, Heisenbergstrasse 3, D-70569 Stuttgart, Germany

[‡]Kirchhoff Institute for Physics, University of Heidelberg, Im Neuenheimer Feld 227, D-69120, Heidelberg, Germany

CONSPECTUS:

The key component of nanoplasmonics is metals. For a long time, gold (Au) and silver (Ag) have been the metals of choice for constructing plasmonic nanodevices, given their excellent optical properties. However, these metals possess a common characteristic, *i.e.*, their optical responses are static. In the past decade, tremendous interest has been witnessed in dynamically controlling the optical properties of plasmonic nanostructures. To enable dynamic functionality, several approaches have been proposed and implemented. For instance, plasmonic nanostructures can be fabricated on stretchable substrates or on programmable templates so that the interactions between the constituent metal nanoparticles and therefore the optical responses of the plasmonic systems can be dynamically changed. Also, plasmonic nanostructures can be embedded in tunable dielectric materials, taking advantage of the sensitive dependence of the localized surface plasmonic resonances on neighboring environment. Another approach, which is probably the most intriguing one, is to directly regulate the carrier densities and dielectric functions of the metals themselves.

In this Account, we discuss a relatively new member in nanoplasmonics, magnesium (Mg), and its important role for the development of dynamic plasmonic nanodevices at visible frequencies. We first elucidate the basic optical properties of Mg and compare it with conventional plasmonic materials such as Au, Ag, and others. Then, we describe a unique characteristic of Mg, *i.e.*, its reversible phase transitions between metallic state and dielectric state, magnesium hydride ($MgH_2$), through hydrogenation and dehydrogenation. This sets the basis of Mg for dynamic nanoplasmonics. In particular, the structural properties and dielectric functions of the two distinct states are discussed in detail. Subsequently, we highlight the experimental investigations on the physical mechanisms and nanoscale understanding of Mg nanoparticles during hydrogenation and dehydrogenation. We then introduce a plethora of newly developed Mg-based dynamic optical nanodevices for applications in plasmonic chirality switching, dynamic color displays with Mg nanoparticles and films as well as dynamic metasurfaces for ultrathin and flat optical elements. We also outline the strategies to enhance the stability, reversibility, and durability of Mg-based nanodevices. Finally, we end this Account by outlining the remaining challenges, possible solutions, and promising applications in the field of Mg-based dynamic nanoplasmonics. We envision that Mg-based dynamic nanoplasmonics will not only provide insights into understanding the catalytic processes of hydrogen diffusion in metals by optical means but also it will open an avenue towards functional plasmonic nanodevices with tailored optical properties for real-world applications.

**Introduction**

The beautiful colors exhibited by metal nanoparticles have been known since medieval times. Artisans exploited the effect to create colorful glass, long before the underlying mechanisms were understood. This phenomenon results from the absorption of sunlight by the metal particles embedded in glass. At a specific wavelength depending on the size, shape, environment of the nanoparticle take place the collective oscillations of the conduction electrons, the *so-called* localized surface plasmon resonance (LSPR).[1,2] LSPRs confine the incident field near the nanoparticle at dimensions much smaller than the operating wavelength. This leads to a strong enhancement of the local fields and allows for manipulation of light below the optical diffraction limit. Such characteristics enable a variety of applications in different disciplines including physics, chemistry, biology, materials science, and others.[2-9]

For a long time, the development of plasmonics has been focused on static systems, whose optical properties are fixed once the structures are fabricated. The concept of active plasmonics or dynamic plasmonics was first proposed in 2004 for controlling signals in a waveguide using nanoscale structural transformations.[10] Since then, the research interest along this direction has flourished. Dynamic plasmonics has taken off as a burgeoning subfield of plasmonics, identifying an inevitable transition of plasmonics from static to dynamic.[11,12] In general, there are two distinct routes to dynamically modulating the optical properties of plasmonic systems. First, tuning the conformations of the plasmonic structures so that the interactions between the constituent metal nanoparticles can be dynamically changed. This is not straightforward for lithographically fabricated samples as the structures are generally restricted on substrates. However, bottom-up approaches such as dynamic DNA nanotechnology provide elegant solutions.[6,13-15] Second, tuning the LSPRs of the individual metal nanoparticles by varying their dielectric surroundings or directly regulating the carrier densities and dielectric functions of the metal particles themselves. For the former, materials that can serve as tunable dielectric surroundings are quite versatile. This includes optically active materials such as photochromic molecules[16], J-aggregates[17], quantum dots[5], and perovskites[18], thermoresponsive materials such as gallium[10], vanadium oxide[19], and germanium antimony telluride[20], electrically-driven materials such as liquid crystals[21] and graphene[22], among others. For the latter, metals that can be regulated directly and meanwhile exhibit excellent plasmonic properties are not very choiceful, especially in the visible spectral range.[23-25]

Magnesium (Mg) is one of the promising candidates, as it exhibits excellent optical properties at high frequencies and can absorb/desorb hydrogen, undergoing reversible transitions between metal and dielectric hydride ($MgH_2$) states.[23,26] This offers great opportunities to design and construct dynamic optical nanodevices at visible frequencies. In this Account, we will elucidate the power of Mg for dynamic nanoplasmonics. More specifically, we will first evaluate the plasmonic and dynamic properties of Mg. We will then discuss the physical mechanisms and nanoscale understanding of Mg nanoparticles during hydrogenation and dehydrogenation. Subsequently, a plethora of newly developed Mg-based dynamic optical nanodevices will be introduced and reviewed. This includes applications in plasmonic chirality switching, dynamic color displays with Mg nanoparticles and films as well as dynamic metasurfaces for ultrathin and flat optical elements. We will also discuss the strategies to enhance the stability, reversibility, and durability of the Mg-based nanodevices. Finally, we will end this Account with a conclusion and outlook.

**Plasmonic properties of Mg**

The dielectric function of bulk metals can be described by the Drude model,[27]

$$\varepsilon(\nu) = \varepsilon_r(\nu) + i\varepsilon_i(\nu) = \varepsilon_\infty - \frac{\nu_p^2}{\nu(\nu+i\gamma)} + \varepsilon_{\text{inter}}(\nu) \qquad (1)$$

where $\varepsilon_r(\nu)$ and $\varepsilon_i(\nu)$ are the real and imaginary parts of the dielectric function. $\varepsilon_\infty$ is the high-frequency limit dielectric constant, $\varepsilon_{\text{inter}}(\nu)$ represents the contribution from interband transitions, $\gamma$ is the damping constant, and $\nu_P$ is the plasma frequency. In the quasi-static regime ($d \ll \lambda$), *i.e.*, the particle size is much smaller than the wavelength of light in the surrounding medium, the condition for resonance, known as Fröhlich condition[27] is fulfilled at $\varepsilon_r = -2$, for a Drude metal sphere located in air. $\nu_f \cong \nu_p/\sqrt{3}$ corresponds to the frequency, at which the LSPR can be excited in the metal sphere. The figure of merit, which characterizes the quality of the excited LSPR, can be written as[24]

$$Q_{\text{LSP}} = \frac{\nu}{\Delta\nu} \approx -\frac{\varepsilon_r(\nu)}{\varepsilon_i(\nu)} \qquad (2)$$

Sanz *et al.* systemically compared the LSPR positions (Fröhlich energy $E_f = h\nu_f$) and values of $Q_{\text{LSP}}^{\max}$ for different metals including Mg, Au, aluminum (Al), Ag, and others as shown in Figure 1.[23] In general, metals with small $\varepsilon_i(\nu)$ present strong and narrow resonances. It is apparent that Mg has excellent plasmonic properties, superior to most of the metals (see Figure 1). More specifically, when compared to conventional plasmonic materials such as Au, Ag, and Al, Mg can produce sharper resonances than Au and Al but is not as good as Ag. Nevertheless, Mg is a more promising material for UV plasmonics than Ag, as interband transitions for Ag already start near 4 eV. In contrary, Mg has no *d*-shell electrons and thus no interband transitions involving *d*-shell electrons occur.[28]

**Dynamic properties of Mg**

Mg reacts reversibly with hydrogen to form magnesium hydride (MgH$_2$),[26,29] which follows the reaction: Mg + H$_2$ ↔ MgH$_2$ + 75.2 kJ mol$^{-1}$. MgH$_2$ is an ionic compound with an appreciable covalent contribution. It exhibits a charge density distribution of Mg$^{+1.91}$ H$^{-0.26}$, in which Mg is almost fully ionized but H is very weakly ionized.[30] The diffusing species in Mg is the H-anion, whose diffusion rate is much slower than the H proton in vanadium (V), niobium, and palladium (Pd).[31]

Mg has a hexagonal-close-packed (hcp) structure with lattice parameters of $a_1 = a_2 = a_3 = 3.21$ Å, $c = 5.21$ Å (see Figure 2 A).[32,33] Previous studies showed that for Mg films and particles deposited on substrates without lattice matching, for instance, on glass, silicon, aluminum oxide (Al$_2$O$_3$), and titanium (Ti), they attained energetically favorable orientations, determined by interface energies, where the facets corresponded to the closest packed planes.[34-36] In other words, the hexagonal prism is the most preferable shape with the Mg [0001] direction perpendicular to the substrate plane.[34] When a small quantity of hydrogen dissolves into the Mg crystal lattice, the α-MgH$_x$ phase (interstitial solid solution of H in Mg) is formed and it has a hexagonal crystal structure.[29,37] With further hydrogenation, β-MgH$_2$ that has a tetragonal rutile crystal structure occurs with lattice parameters of $a = b$ = 4.52 Å, $c$ = 3.02 Å (see Figure 2A).[32,33] In between these two phases, there exists a mixed region, the α+β phase, in which hydrogen dissolved in Mg is in equilibrium with MgH$_2$. The plateau pressure for the transition from the α-phase to the α+ β phase is 0.41 Torr at 353K.[29] Electron diffraction studies revealed that the most intrinsic orientation relationship between Mg and MgH$_2$ lattices during the phase transition was Mg(0001)[2$\bar{1}\bar{1}$0] ∥ MgH$_2$(110)[001] (see insets in Figure 2A).[32,33] Due to the atomic movements resulting from the phase transition, the distance between the Mg atoms is expanded by 23% (from $c_{Mg}$ to $\sqrt{2}a_{MgH_2}$) along the direction perpendicular to the substrate plane. The distance between the Mg atoms in the substrate plane is expanded by only 6%. Hence, major lattice distortions occur along the out-of-plan direction.[38]

Mg is a potential material for solid-state hydrogen storage owing to its abundance, low cost, reversibility, large hydrogen gravimetric (7.6 wt %), and volumetric (110 g l$^{-1}$) capacities.[26] However, there are two major obstacles for practical applications: high hydrogenation/dehydrogenation temperatures and sluggish hydrogen absorption/desorption kinetics. The surface of pure Mg has a large activation energy for hydrogen dissociation and hydride formation. Hydrogenation of Mg requires high operating temperatures up to ~300 °C at 1 atm pressure and dehydrogenation needs even higher temperatures ~400 °C.[39,40] It was discovered that Pd capping on Mg could reduce the high operating temperatures to ambient conditions by catalyzing the dissociation of H$_2$ molecules.[40] In this case (see Figure 2B), as the reaction progresses MgH$_2$ grows at the Mg/Pd interface. The kinetic limiting step is hydrogen diffusion through the growing hydride layer MgH$_2$, which acts as a barrier for further hydrogenation of Mg. This is the *so-called* blocking effects.[41] During dehydrogenation, the growing metallic Mg near the Mg/Pd interface then limits the desorption kinetics (see Figure 2C).

MgH$_2$ is a transparent and color neutral insulator with a band gap of 5.6 ± 0.1 eV.[42] MgH$_2$ can be regarded as a nearly non-dispersive and low-loss dielectric material with refractive index $n+ik$ = 1.95 + $i$0.01 in the visible and near-infrared regimes.[42] Due to the large contrast between Mg and MgH$_2$ both in optical and electrical properties, the transition processes can be conveniently investigated by joint optical and electrical measurements in real time.[43]

**Dynamic plasmonic nanoparticles**

In 1997, van der Sluis *et al*. demonstrated optical switching of Mg-based films between mirror and transparent states through hydrogenation and dehydrogenation.[44] Since then, 'switchable mirrors' have been widely investigated. This has led to a variety of applications including optical hydrogen sensors[45], switchable solar absorbers[46], and smart windows.[47] However, the blocking effects as a bottleneck problem remained, hampering the further development for practical applications. Later, Uchida *et al.* showed that Mg films exhibited favorable absorption kinetics only for film thickness below ~100 nm with diffusion coefficient of ~$10^{-16}$ m$^2$s$^{-1}$.[37] Subsequently, a MgH$_2$ layer was formed at the Mg/Pd interface and the diffusion coefficient decreased to ~$10^{-18}$ m$^2$s$^{-1}$, preventing hydrogen from effective diffusion. Although the kinetics could be improved by increasing temperature and/or decreasing hydrogen concentration,[41] the formation of MgH$_2$ as diffusion barrier dramatically slowed down the diffusion process, especially when the diffusion length was longer than ~100 nm.

To enhance diffusion kinetics, Mg and MgH$_2$ particles have been utilized.[48] Small particles possess higher surface-to-volume ratios than large particles, giving rise to faster reaction rates. Decreasing the crystal grain size can also reduce the thermodynamic stabilities of Mg and MgH$_2$, resulting in lower hydrogen absorption and desorption temperatures.[49] The most common method to producing Mg particles is mechanical milling. However, the lack of quantitative control over the sizes, morphologies and compositions has encumbered the understanding of hydrogenation/dehydrogenation mechanisms on the nanoscale.

In recent years, advances in nanofabrication techniques have offered great opportunities to pattern Mg nanoparticles with controlled sizes, shapes, compositions, and surface morphologies. This has also enabled a wealth of Mg-based plasmonic nanostructures. For instance, Sterl *et al.* fabricated Mg nanoparticles of various sizes by colloidal hole-mask lithography and subsequent electron-beam evaporation (see Figure 3A).[50] These Mg particles possessed hexagonal monocrystalline shapes, when their sizes were relatively small (~100 nm). They tended to be polycrystalline with increasing sizes. These Mg nanoparticles also exhibited pronounced LSPRs, which were tunable throughout the visible wavelength range by varying the particle size. With the help of Pd as catalytic layer, the resonances could be turned off/on upon hydrogen/oxygen exposure at room temperature, when the particles were transformed between the metallic Mg state and the dielectric MgH$_2$ state (see Figure 3B). The weak resonances still observable in the MgH$_2$ state mainly resulted from the incomplete hydrogenation of the Mg particles. To avoid the Mg-Pd alloy formation[39], 5 nm Ti was used to separate Mg and Pd. This buffer layer also helped to release the mechanical stress, resulting from the different expansion rates of Mg and Pd upon hydrogen absorption.

**Nanoscale hydrogenography on single Mg nanoparticles**

The spatial resolution of conventional hydrogenography in the horizontal plane is optical diffraction limited. This prevents direct optical investigations of hydrogen diffusion in individual Mg crystallites, which are typically on the size scale of 100 nm. To understand hydrogenation of Mg particles on the nanoscale, Sterl *et al.* utilized scattering-type scanning near-field optical microscopy (s-SNOM), which could locally probe the dielectric properties of matter with spatial resolution on the order of tens of nanometers.[51] Dark-field spectroscopy was employed to measure the scattering spectra of single Mg particles, which were used to evaluate the amount of metallic Mg within each particle.

The s-SNOM images recorded after different $H_2$ exposure durations (see Figure 3D) revealed that the hydrogenation process was inhomogeneous both temporally and spatially in the Mg nanoparticles. The phase transition from Mg to $MgH_2$ was rapid within a single crystallite before progressing towards adjacent ones. Each particle exhibited an individual hold-back time in the beginning. This could be attributed to the grain boundaries of the individual crystallites, which acted as barriers for hydrogen diffusion in a single nanostructure.[52] This work proved that the crystalline structure of Mg nanoparticles was crucial for hydrogen absorption and desorption kinetics, providing insights into design and fabrication of Mg-based dynamic optical nanodevices.

**Plasmonic chirality tuning**

Chirality is a geometrical property of an object. A chiral object and its mirror image are called enantiomers, which cannot be superimposed with one another. Apart from geometrical properties, chirality can also manifest itself optically via a different response to left- and right-handed circularly polarized (LCP and RCP) light. The resulting absorption difference is called circular dichroism (CD).[53] In general, CD of natural chiral molecules such as amino acids, proteins, carbohydrates, *etc*., is very weak and located only in the UV spectral region.[53] In contrast, chiral plasmonic structures can exhibit spectrally tunable and pronounced CD, which is several orders of magnitude larger than that of natural chiral molecules.[54-60]

We demonstrated a hydrogen-regulated chiral plasmonic system as shown in Figure 3E.[61] Each chiral structure consisted of four Au and four Mg particles that were arranged in a gammadion-like geometry. On top of the Mg particles, 5 nm Ti and 10 nm Pd were capped for facilitating hydrogen loading/unloading at room temperature. The samples were fabricated using a double electron-beam lithography (EBL) process. Its scanning electron microscopy (SEM) image is shown in Figure 3F.

Before hydrogen loading, the left-handed sample exhibited a bisignate spectral profile as characterized by the red line in Figure 3F. This resulted from the resonant coupling between the collective plasmons excited in the eight plasmonic particles.[62] Upon hydrogen loading, Pd catalyzed the dissociation of hydrogen molecules into atoms, which could diffuse through the Ti spacer into the Mg particles. As a result, the four Mg particles were gradually hydrogenated into $MgH_2$ particles. This was reflected by the successively decreasing CD strength. In the end, the entire structure became achiral, giving rise to a featureless CD spectrum. The chiral spectra could be recovered through dehydrogenation by exposing the sample in ambient air or oxygen. Hence, the chiroptical responses of the plasmonic structures could be dynamically switched off/on simply by hydrogenation/dehydrogenation. Such a dynamic control concept may lead to plasmonic chiral platforms for a variety of gas detection schemes by exploiting the high sensitivity of CD spectroscopy. It is noteworthy that in ambient air Mg can form MgO with oxygen, magnesium hydroxycarbonate with carbon dioxide, and possibly becomes hydroxylated due to the formation of $Mg(OH)_2$. This results in low system reversibility to only several cycles. It can be improved by carrying out the experiment in a dry environment and/or covering a thin polytetra-fluoroethylene protection layer on the Mg surface.

**Dynamic plasmonic color displays based on Mg nanoparticles**

Plasmonic colour generation based on engineered metasurfaces has stimulated a variety of fascinating applications in colour display science for high-density optical data storage, information anticounterfeiting, and data encryption.[7,63] Based on catalytic Mg metasurfaces as shown in Figure 4A, we demonstrated a dynamic plasmonic display technique, which enabled plasmonic microprint displays with good reversibility.[64]

The plasmonic pixels comprising Mg nanoparticles were sandwiched between Ti/Pd capping layers and a Ti buffer layer. On the metasurface, these Mg particles were arranged in a lattice with various particle sizes and interparticle distances to achieve brilliant colors in a broad range (see Figure 4B). Through hydrogenation, different color squares underwent a series of vivid color changes until all colors vanished. The hydrogenation process was essentially associated with a gradual decrease of the metallic fraction of the particles, forming $MgH_2$ as dielectric surrounding. Such a catalytic process rendered dynamic alterations to the reflectance spectra and therefore the exhibited colors possible.

A plasmonic microprint was fabricated based on the Max-Planck-Society's Minerva logo as shown in Figure 4C. Upon hydrogen loading, the Minerva logo experienced dynamic color changes. Abrupt color alterations took place within 23 s. Subsequently, the logo started to fade and completely vanished after 566 s. Upon oxygen exposure, the logo could be restored to its starting state. We further showed that through smart material processing, information encoded on selected pixels, which were indiscernible to both optical and scanning electron microscopies, could be read out using hydrogen as a decoding key, suggesting a new generation of information encryption and anti-counterfeiting applications.

**Dynamic color displays based on Mg cavity resonances**

To simplify fabrication procedures and enhance the durability of Mg-based dynamic displays, we reported a dynamic color display scheme using pixelated Fabry-Pérot (FP) cavities (see Figure 5A).[65] Each pixel (500 × 500 $nm^2$) consisted of a dielectric hydrogen silsesquioxane (HSQ) pillar, which was sandwiched between an Al mirror and a metallic capping layer composed of Mg/Ti/Pd (50 nm/2 nm/3 nm). By tuning the pillar heights using grey-scale lithography, a series of FP cavities were formed.

Before hydrogenation, the Mg/Ti/Pd capping layer efficiently reflected the visible light, resulting in no color generation. This defined a blank state (see Figure 5B). Upon hydrogen exposure, Mg was gradually transformed to $MgH_2$, the effective thickness of the metallic capping layer decreased and light started to transmit through it. When Mg was fully hydrogenated into $MgH_2$, colors were selectively reflected from these FP resonators of different heights. In this case, each FP resonator consisted of a $TiH_2$/PdH capping layer, a double dielectric spacer ($MgH_2$ + HSQ), and an Al back mirror. Such asymmetrical FP resonators with ultrathin lossy capping generated vivid and high-contrast colors with a wide gamut, representing a color state. The resonance properties such as the reflectance peak positions and the number of allowed modes in the FP resonators were largely governed by the individual cavity heights as shown in Figure 5B. This scheme utilized a Mg layer directly from thin film deposition to achieve dynamic color changes without any post-nanofabrication steps.

**Scanning plasmonic color displays**

Very recently, we demonstrated a novel scanning plasmonic color display, taking inspiration from macroscopic scanning devices.[66] As shown in Figure 5C, the microscopic scanning screen was a Ti/Mg (5nm/30 nm) layer with dimensions of 15 × 15 μm$^2$ on a SiO$_2$ substrate. To enable the scanning characteristics, the left side of the screen was in contact with a Pd strip, which worked as a gate for hydrogen loading or unloading. Al nanoparticles, *i.e.*, the plasmonic pixels, were arranged on top of the scanning screen spaced by a 20 nm Al$_2$O$_3$ layer. Upon hydrogen loading, hydrogenation of Mg started from the Pd gate such that the plasmonic pixels were laterally scanned, following the hydrogen diffusion direction. During the process, the scanning screen transited from a mirror (Mg) to a transparent spacer (MgH$_2$). This process was reversible through dehydrogenation using oxygen.

In particular, we carefully investigated lateral hydrogen diffusion in Mg,[67] which had been rarely studied before. In contrast to the vertical diffusion scheme (*i.e.*, out-of-plane diffusion),[26,29,37] we discovered that the blocking effects were absent in a long-range lateral diffusion over tens of micrometers. In addition, the lateral diffusion was fast at all times and not hampered by the MgH$_2$ barrier layer. In order to characterize the diffusion front mobility, optical hydrogenography was utilized to *in situ* record the optical reflection images of the diffusion process. As shown in Figure 5D, the experimental data revealed a typical diffusive process following a nucleation step. The square of the front position $x^2$ was proportional to time $t$, after a short nucleation time, $t_0$. The front mobility $K$ could be obtained by fitting the experimental curve. Similarly, the diffusion parameters associated with the dehydrogenation process could be experimentally obtained as well.

**Mg-based dynamic metasurfaces**

Other than plasmonic color generation, in which only the amplitude of light is tailored, metasurfaces can also manipulate the phase of light at an unprecedented level. This capability has enabled a wealth of ultrathin optical devices for beam focusing and steering[8], vortex beam generation[68], and holography[69].

We demonstrated an Mg-based dynamic metasurface platform, which allowed for independent manipulation of addressable subwavelength pixels at visible frequencies through controlled hydrogenation and dehydrogenation.[70] Such plasmonic pixels, consisting of Mg nanorods with various orientation angles, were utilized to control the light wavefronts via the Pancharatnam-Berry (PB) phase.[69] To shape arbitrary light wavefronts, eight-phase levels were chosen for the metasurfaces as shown in Figure 6A. To achieve holographic patterns with sequenced dynamics, we multiplexed the metasurface with dynamic pixels that possessed different reaction kinetics upon hydrogenation/dehydrogenation. As shown by the SEM image in Figure 6B, each unit cell contained a Mg/Pd (P$_1$) nanorod and a Mg/Pd/Cr (P$_3$) nanorod as two sets of dynamic pixels. The Cr (1 nm) capping layer could effectively slow down both the hydrogenation and dehydrogenation rates of $P_3$. This led to distinct time evolutions of $P_1$ (solid line) and $P_3$ (dash-dotted line) during hydrogenation and dehydrogenation (see Figure 6C). Two holographic patterns were reconstructed based on $P_1$ and $P_3$ using RCP light. As shown in Figure 6D, the portrait of Marie Curie as well as the chemical elements Po and Ra, could transit among four distinct states through hydrogenation and dehydrogenation.

Taking a step further, we demonstrated dynamic Janus metasurfaces at visible frequencies.[71] Each super unit cell comprised three pixels that were categorized in two sets (see Figure 6E). In one set, a Au nanorod ($P_+$) and a Mg nanorod ($P_-$) were orthogonally arranged as counter pixels. The anomalous RCP waves reflected from $P_+$ and $P_-$ achieved a $\pi$ phase shift, giving rise to a reflectance minimum in the far field due to destructive interference. In the other set, there was an additional Mg nanorod (P). Before hydrogenation, the effective pixel of the super unit cell was $P_e = P$. After hydrogenation, Mg was transformed into $MgH_2$. The net function of the super unit cell was therefore only governed by $P_+$, *i.e.*, $P_e = P_+$. Upon oxygen exposure, $MgH_2$ could be transformed back to Mg and thus $P_e = P$ again. Therefore, the effective pixels on such a Janus metasurface could be reversibly regulated using $H_2$ and $O_2$, independent on the helicity of the incident light.

This scheme opened a unique pathway to endow optical metasurfaces with rich dynamic functionalities in the optical spectral region, especially for generation of metasurface holography with high security. As illustrated in Figure 6F, before hydrogenation 'Y' was observed on the screen upon illumination of RCP light. After $H_2$ exposure, 'Y' vanished, whereas 'N' came into existence. After $O_2$ exposure, 'N' disappeared and 'Y' occurred again. Consequently, the two holographic images, containing different information could be independently reconstructed upon $H_2$ and $O_2$ exposures, respectively, while the helicity of the incident light remained unchanged.

**Stability, reversibility, and durability**

Mg has higher chemical reactivity than widely used plasmonic materials such as Au, Ag, and Al. Concerns about stability, reversibility, and durability as well as approaches to solve these issues are highly relevant. The corrosion kinetics of Mg under ambient conditions is mainly governed by two factors: the water content in air and the characteristics of the Mg surface.[72] The corrosion of Mg happens rapidly in humid air. Nevertheless, it was observed that Mg films and particles were very stable, when exposed to dry air owing to the formation of very thin MgO passivation shells. In humid environment, MgO may absorb moisture to form $Mg(OH)_2$, which no longer serves as a protection layer for Mg. This hydration process of MgO is strongly influenced by the crystallographic orientation and the presence of defects on the oxide surface. Kooi *et al*. demonstrated that crystallized Mg nanoparticles exhibited a dense and crystalline MgO shell (~3 nm), preventing further oxidation under ambient conditions at room temperature for one month.[34] In contrast, poorly crystallized Mg with low-density surface possessed a porous and amorphous MgO shell, which could hydride easily to form a porous $Mg(OH)_2$ layer. Therefore, improving the crystallinity of Mg during fabrications and operating the Mg-based nanodevices in dry environment are critical actions to consider. To operate Mg-based nanodevices in humid air, a thin polytetrafluoroethylene layer can be deposited on the Mg surface to isolate water but still allow for hydrogen diffusion.[73] In addition, for all the aforementioned Mg-based nanodevices, the utilization of Pd and Ti capping layers was proved to be very effective for Mg protection, showing good device performance in terms of reversibility and durability.

**Conclusion and outlook**

Mg for dynamic nanoplasmonics is a viable route to realizing plasmonic nanodevices with novel functionalities, given its design flexibility and large modulation of the optical responses. There are remaining issues, which need to be addressed for construction of high-performance dynamic systems for real-world applications.

First, optical approaches, such as s-SNOM provide physical understanding of the *in-situ* hydrogenation and dehydrogenation processes with resolution of several tens of nanometers. Deeper insights on such processes on the atomic level can be achieved using environmental transmission electron microscopy (TEM).[74,75] For instance, Mg nanoparticles of different sizes, shapes, geometries, *etc.*, can be fabricated by advanced EBL on an ultra-thin TEM grid. Tomography and diffraction patterns of the particles before and after hydrogenation can be obtained and carefully examined. In addition, *in-situ* electron-energy loss spectropy of the Mg nanoparticles can be carried out during hydrogenation/dehydrogenation. Such characterizations will enable visualization of the phase-transition dynamics of the Mg nanoparticles in a controlled gaseous environment and allow for understanding the fundamental atomic mechanisms of gas-solid reactions on the atomic level. In turn, the gained knowledge of the time-resolved dynamic and kinetic mechanisms on the atomic level will provide insightful blueprints to design Mg-based dynamic nanodevices with high performance. Second, to further improve reversibility and durability, alloying Mg with other metals including nickel (Ni), yttrium (Y), V, iron (Fe), *etc.*, should be attempted.[26] Kalisvaart *et al.* showed that Mg films alloyed with Al, Fe, and Ti could enhance reaction kinetics and no degradation in performance was observed after 100 absorption/desorption cycles.[76] Baldi *et al.* and Slaman *et al.* demonstrated that Pd capped Mg-Ti alloy films showed faster kinetics with good reversibility over 150 cycles.[45,46] Remarkably, Tajima *et al.* reported optical switching of $Mg_4Ni$ films over 4000 cycles.[47] Therefore, research efforts on patterning Mg alloy particles for dynamic optical nanodevices will be very rewarding to improve the device reversibility and switching rates. Third, to enhance switching rates, other hydrogenation/dehydrogenation means can be considered. Den Broeder *et al.* demonstrated electromigration of hydrogen in Y films, in which the diffusing species in the insulating yttrium trihydride was the H-anion.[77] This concept can be applied to Mg-based systems as well so that the hydrogenation/dehydrogenation processes in Mg can be conveniently controlled by external electric fields at fast speeds. This will also eventually solve the portability issue of the gas-phase reactions.[47]

So far, Mg-based dynamic nanodevices have been utilized for applications in plasmonic chirality switching, dynamic color displays and metasurface elements. This exploits only a glimpse of opportunities that Mg can offer. There are many interesting topics that deserve research endeavors. For instance, the out-of-plane expansion of Mg nanoparticles is as large as 30%.[38,51] This provides a unique model system to realize plasmonic devices with nanoscale mechanical responses. Also, Mg can be utilized in the studies of switchable nonlinear effects, offering a tailored platform to examine the intriguing enhancement and symmetry questions in nonlinear plasmonics. Furthermore, Mg can be applied for plasmonic sensing, which is not only limited to detection of hydrogen. Rather, it can be extended to offer general dynamic platforms for tunable surface-enhanced Raman scattering, fluorescence, infrared absorption, and others. We believe the unsolved challenges and new scientific inquires will stimulate exciting and continuous studies of Mg-based dynamic nanoplasmonics and their related applications.

AUTHOR INFORMATION

Corresponding Author

*E-mail: na.liu@kip.uni-heidelberg.de

ORCID

Na Liu: 0000-0001-5831-3382
Xiaoyang Duan: 0000-0002-8720-3788

Author Contributions

The manuscript was written by both authors. Both authors have given approval to the final version of the manuscript.

Notes

The authors declare no competing financial interest.

Biographical Information

Xiaoyang Duan is a Ph.D. candidate in Prof. Liu's group at the University of Heidelberg and the Max Planck Institute for Intelligent Systems, Germany.

Na Liu is a Professor at the Kirchhoff Institute for Physics at the University of Heidelberg, Germany.

ACKNOWLEDGMENTS

We acknowledge support from the European Research Council (ERC Dynamic Nano) grant.

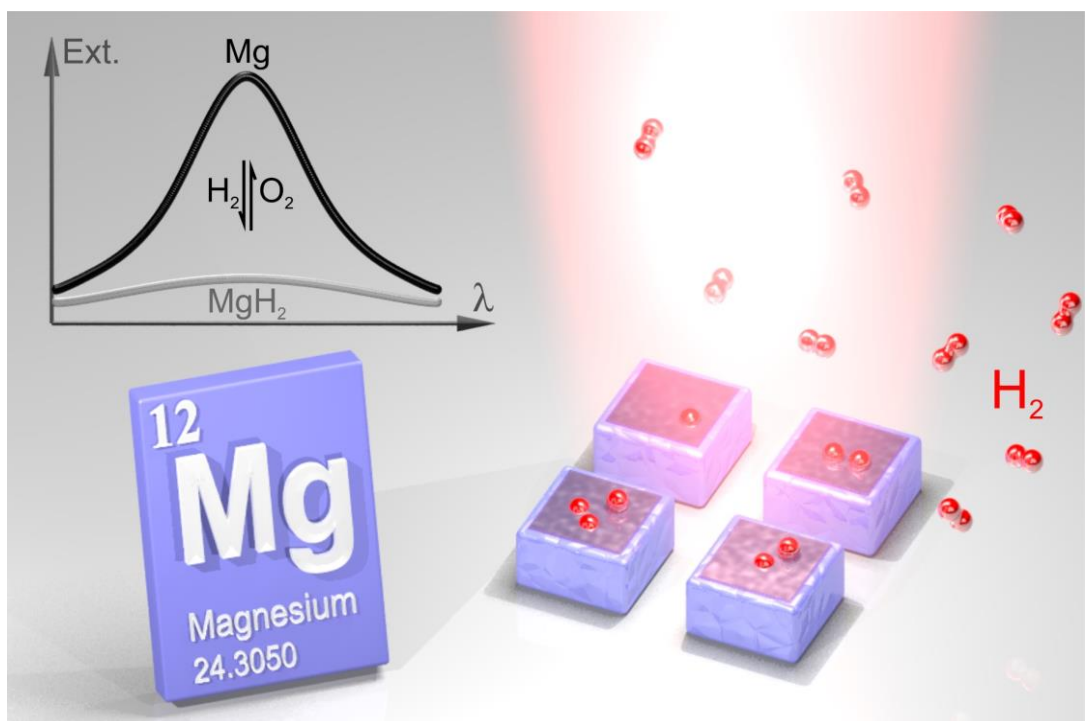

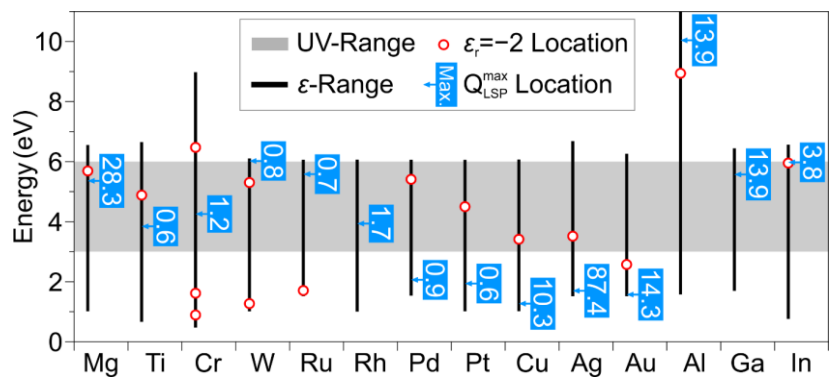

**Figure 1.** Evaluation of the plasmonic properties of different metals. Fröhlich energy and the maximum plasmonic performance value, $Q_{\text{LSP}}^{\max}$, plotted at the spectral position, where it is achieved. Reproduced with permission from ref 23. Copyright 2013, American Chemical Society.

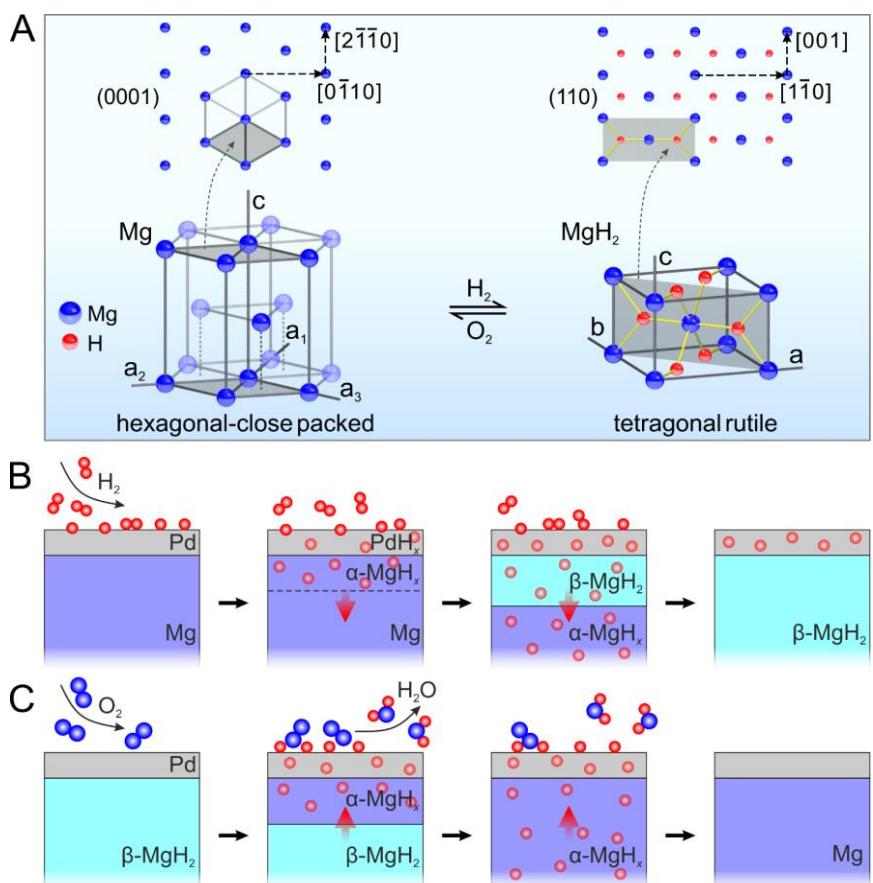

**Figure 2.** (A) Crystallographic phase transformations between Mg and MgH$_2$. Insets: atomic arrangements in the Mg (0001) and MgH$_2$ (110) planes, respectively, which are parallel to the substrate. Schematic models of hydrogenation (B) and dehydrogenation (C) processes for the Pd-capped Mg upon hydrogen and oxygen exposure, respectively.

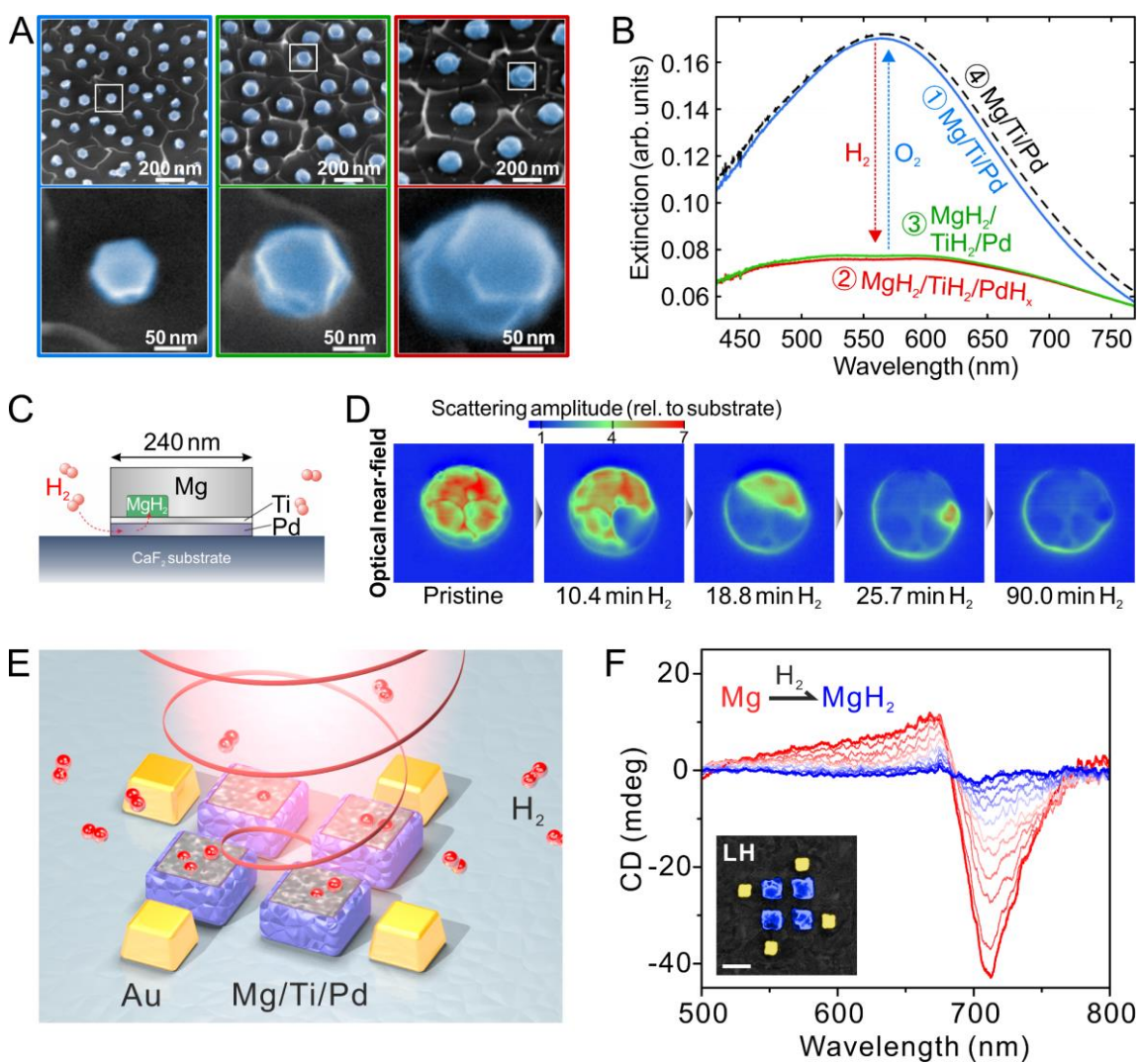

**Figure 3.** (A) Colorized SEM images of Mg nanodisks with different diameters. (B) Extinction spectra of the 80 nm Mg/5 nm Ti/10 nm Pd particles with diameter of 160 nm in the different stages of a typical hydrogenation/dehydrogenation cycle. (C) Schematic of the Mg nanostructure, consisting of 40 nm Mg/5 nm Ti/10 nm Pd. (D) Near-field scattering maps of the Mg particle recorded between hydrogen exposures, covering an area of 600 × 600 nm$^2$. (E) Schematic of a hybrid chiral plasmonic system, which consists of four Mg/Ti/Pd particles and four Au particles in a gammadion-like arrangement. (F) Evolution of the measured CD spectra upon hydrogen loading as a function of time. Scale bar: 200 nm. Reproduced with permission from (A, B) ref 50, (C, D) ref 51, and (E, F) ref 61. Copyright 2015, 2018, and 2016, respectively, American Chemical Society.

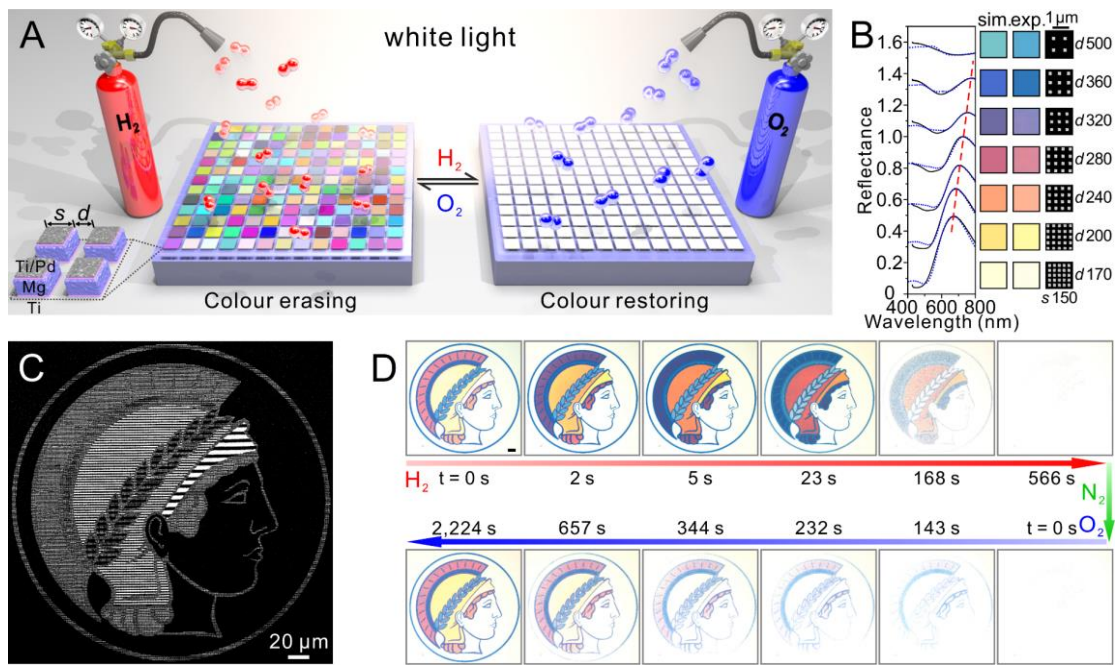

**Figure 4.** (A) Schematic of the plasmonic metasurface composed of Mg nanoparticles. (B) Experimental (black) and simulated (blue-dotted) reflectance spectra and colors as well as the corresponding SEM images of the structures. (C) Overview SEM image of the Minerva logo sample. (D) Optical micrographs of the Minerva logo during hydrogenation and dehydrogenation for color tuning, erasing and restoring. Scale bar: 20 μm. Reproduced with permission from ref 64. Copyright 2017 Nature Publishing Group.

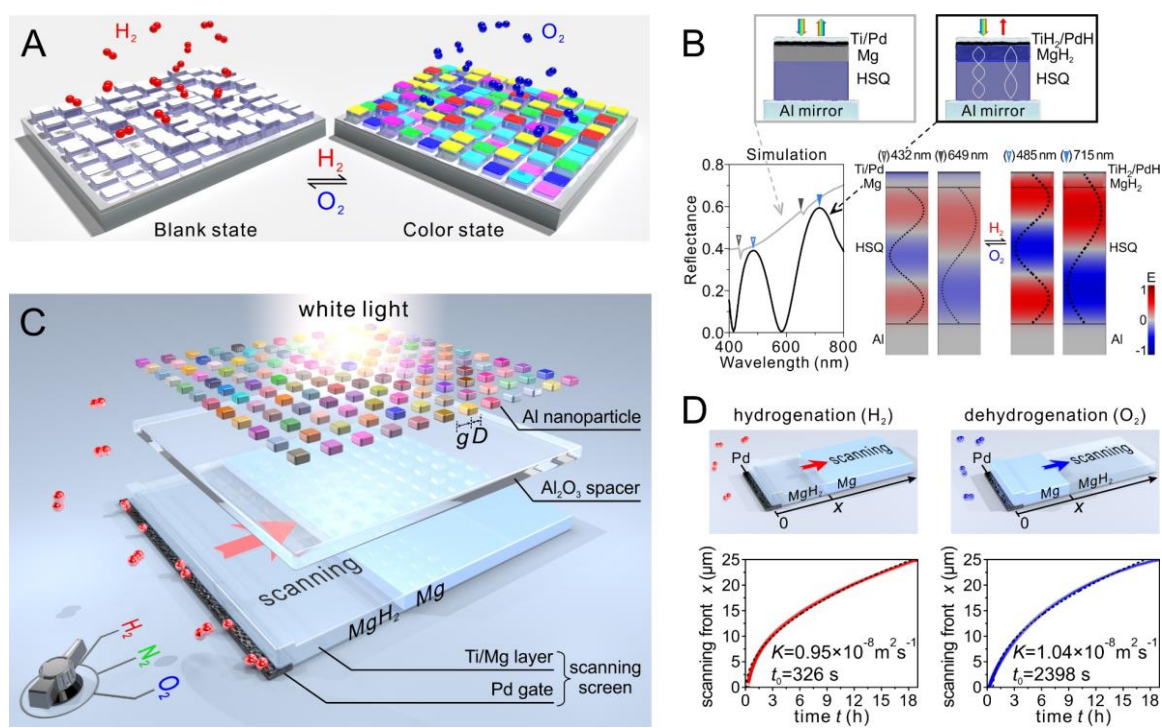

**Figure 5.** (A) Schematic of the dynamic color display using stepwise FP resonators. Pixelated HSQ pillars of different heights generated by gray scale nanolithography are sandwiched between a Mg/Ti/Pd (50 nm/2 nm/3 nm) capping layer and an Al mirror. (B) Simulated reflectance spectra of a representative FP resonator in the blank (grey line) and color (black line) states, respectively. The electric field distributions of the different FP resonances (highlighted using arrows) before and after hydrogenation. (C) Schematic of the scanning plasmonic color display, which consists of Al nanoparticles as plasmonic pixels, a 20 nm dielectric $Al_2O_3$ spacer, and a scanning Mg screen (15 μm × 15 μm × 30 nm) with a 3 nm Ti buffer layer. (D) Scanning front $x$ of the Mg screen during hydrogenation (red) and dehydrogenation (blue) at different times tracked by *in situ* optical hydrogenography. Reproduced with permission from (A, B) ref 65, and (C, D) ref 66. Copyright 2017, and 2018, respectively, American Chemical Society.

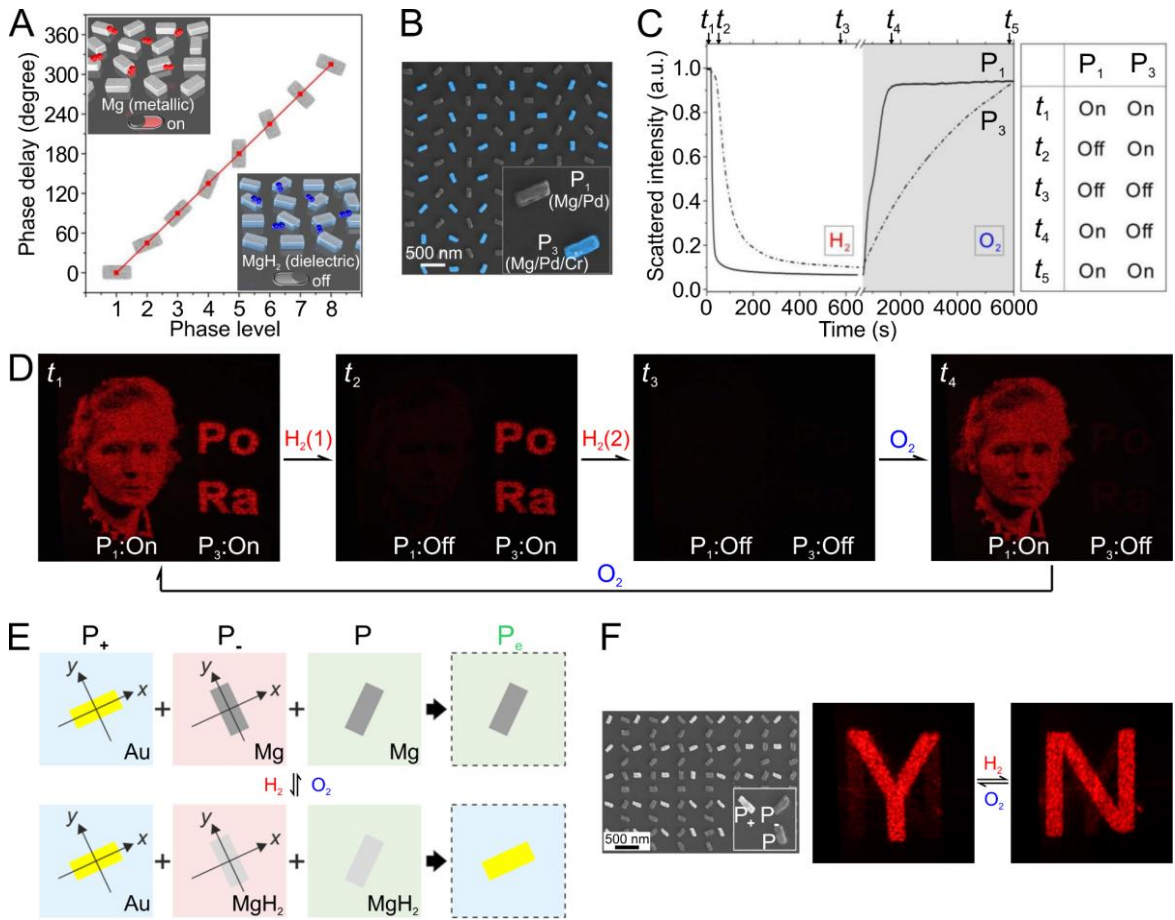

**Figure 6.** (A) Schematic of the metasurface hologram consisting of Mg nanorods with different orientations and the simulated phase delay with respect to the orientation. (B) Overview SEM image of the hybrid metasurface. (C) Evolution of the scattered intensities of $P_1$ (solid line) and $P_3$ (dash-dotted line) during hydrogenation and dehydrogenation. (D) Representative snapshots of the holographic images during hydrogenation and dehydrogenation. (E) Working principle of the dynamic Janus metasurface. (F) Snapshots of the holographic images before and after hydrogenation. (A–C) Reproduced with permission from ref 70. Copyright 2018 AAAS. (E, F) Reproduced from ref 71. Copyright 2018 American Chemical Society.